\documentclass[trackchanges,twocolumn]{aastex631}

\newcommand{\V}[1]{\mathbf{#1}} %Bold for Vectors
\newcommand\pplume{$\mathcal{W}$ } %Proper Names

\usepackage[edges]{forest}
\forestset{
  colour my roots/.style={
    before typesetting nodes={
      edge=#1,
      for ancestors={
        edge=#1,
        #1,
      },
      #1,
    }
  },
  my edge label/.style={
    edge label={
      node [midway, fill=white, font=\scriptsize] {#1}
    }
  }
}

\newcommand{\uofa}{\affiliation{Lunar and Planetary Laboratory, University of Arizona, Tucson, AZ 85721, USA.}}
\newcommand{\lesia}{\affiliation{LESIA, Observatoire de Paris, Universit\'e PSL, CNRS, Sorbonne Universit\'e, Universit\'e de Paris, 92195 Meudon, France.}}

\received{Feb 25, 2023}
\revised{Apr 23, 2023}
\accepted{June 3, 2023}

\shorttitle{\title{Classification of Ion-Driven Instabilities}}
\shortauthors{Martinovi\'c \& Klein}

\begin{document}

\title{Ion-Driven Instabilities in the Inner Heliosphere II: Classification and Multi-Dimensional Mapping }

\correspondingauthor{M. M. Martinovi\'c}
\email{mmartinovic@arizona.edu}

\author[0000-0002-7365-0472]{Mihailo M. Martinovi\'c}
\uofa
\lesia

\author[0000-0001-6038-1923]{Kristopher G. Klein}
\uofa

%%%%%%%%%%%%%%%%
%%% ABSTRACT %%%
%%%%%%%%%%%%%%%%

\begin{abstract}

Linear theory is a well developed framework for characterizing instabilities in weakly collisional plasmas, such as the solar wind. 
In the previous instalment of this series, we analyzed $\sim1.5$M proton and $\alpha$ particle Velocity Distribution Functions (VDFs) observed by \emph{Helios I} and \emph{II} to determine the statistical properties of the standard instability parameters such as the growth rate, frequency, the direction of wave propagation, and the power emitted or absorbed by each component, as well as to characterize their behavior with respect to the distance from the Sun and collisional processing. 
In this work, we use this comprehensive set of instability calculations to train a Machine Learning algorithm consisting of three interlaced components that: 1) predict if an interval is unstable from observed VDF parameters; 2) predict the instability properties for a given unstable VDF; and 3) classify the type of the unstable mode. 
We use these methods to map the properties in multi-dimensional phase space to find that the parallel-propagating, proton-core-induced Ion Cyclotron mode dominates the young solar wind, while the oblique Fast Magnetosonic mode regulates the proton beam drift in the collisionally old plasma. 
\end{abstract}


%%%%%%%%%%%%%%%%%%%%
%%% INTRODUCTION %%%
%%%%%%%%%%%%%%%%%%%%

\section{Introduction} \label{sec:intro}

Solar wind plasma is rarely observed to be in Local Thermodynamic Equilibrium (LTE), but rather contains non-Maxwellian features that imply additional free energy stored in the constituent particles' Velocity Distribution Function (VDF) (for review, see \citep{Marsch_2012_SSRv,Verscharen_2019_LRSP}). 
If the VDF is not far from equilibrium, rare collisions constantly reshaping it towards a Maxwellian through a slow and steady process. 
However, if the distribution is sufficiently far from LTE, it will drive one or more unstable wave modes, where power is emitted from the particles in the form of waves.
This emission occurs over significantly shorter timescales then the collisional processing, pushing the VDF into a state known as "marginal stability"---where no further instabilities are induced, but the distribution is not in LTE and the VDF continues to be slowly processed by collisions. 
Although linear instabilities, as well as prescriptions for their identification, are well established in the literature \citep{Gary_1993,Klein_2013_PhD,Yoon_2017}, a detailed description of which modes govern solar wind evolution through its various phases as it expands into the heliosphere, and how instabilities and collisions interact, is still incomplete. 

The preceding paper in the series \citep{Martinovic_2021_ApJ_Ins_1} (further on referred to as "Paper I"), provided a statistical analysis of the instability occurrence rate and nature of predicted waves by analyzing VDF data sampled by \emph{Helios I} and \emph{II} between 0.3 and 1 au and fitted as a sum of Maxwellian componenents \citep{Durovcova_2019_SoPh}. 
Processing $\sim1.5$M VDFs using the Plasma in a Linear Uniform Magnetized Environment (\texttt{PLUME}) dispersion solver \citep{Klein_2015_PhPl} created a rich data set of $\sim630$K unstable intervals. 
Organizing the results by different solar wind parameters, we concluded that the Coulomb number---the estimated number of Coulomb thermalization times $N_{C(cc)} = \nu_{cc} r / v_{sw,c}$, with $\nu_{cc}$ the collision frequency of core protons, see \cite{Hernandez_1987,Kasper_2017} for details--- is the strongest indicator of both how often unstable modes are induced and the amplitude of their growth rates.
In the young solar wind, over 80\% of intervals were found to be unstable. 
As the collisional processing becomes significant, that percentage declines exponentially, until we reach collisionally old wind close to LTE, where instabilities are predicted to arise less than 10\% of the time, and the associated growth rates are significantly weaker than those encountered in younger wind.

A natural expansion of this result would be to provide a more complete picture of the relation between various instability characteristics and VDF parameters of interest for the solar wind and heliospheric plasmas. 
This task has turned out to be very complicated, primarily because of two major issues: 
(1) even though the data set is very large, some parts of the phase space are filled rather sparsely due to instrument limitations or features of the fitting algorithm (both discussed in detail in Paper I), driving a need for prediction of the unstable mode properties for a generic VDF, and; 
(2) difficulty of automatized identification and classification of any given unstable mode. 
For a given interval, this task is straight-forward, and sometimes fairly simple. 
However, building an automatized process that takes into account 11 or more features of the unstable mode, e.g. frequency, growth rate, direction of propagation, plus up to 13 variables that characterize the VDF, e.g. thermal-to-magnetic pressure ratios, temperature anisotropies and disequilibrium, using statistical methods only was not feasible. 

The main focus of this work is to provide the tools necessary to address these two issues, which is done through the development of customized Machine Learning (ML) models, trained and tested using the processed \emph{Helios} observations. 
In Section \ref{ssec:prediction}, we train the classifier to distinguish if a given VDF is either stable or unstable. 
In Section \ref{ssec:quantifyication}, we describe the regression codes that estimate the behavior of the most unstable mode for an observed particle VDF.
Combination of the two algorithms provides the ability to predict unstable modes for any given VDF---widening our research scope to generic distributions represented as a set of Maxwellians, not just the ones observed---addressing problem (1) for the parameter range of interest. 
To resolve (2), we build a classification algorithm that divides the unstable modes into clusters based on their weighted distance in phase space (Section \ref{ssec:classification}). 
The parameters of the clusters correspond very well with characteristics of theoretically described modes, enabling a physical interpretations of the kinds of modes expected to be driven unstable. 
This feature is the main distinction between traditional dispersion solvers and our code. 
Although various instability types can have fundamentally different physical processes as a cause, the numerical parameters of instabilities can be very similar, e.g. similar growth rates, polarizations, or wavevector regions where they are most unstable. 
Therefore, a set of statistical criteria that distinguishes the type of any given mode has not previously been developed. 
Given the complexity of the parameter phase-space, we were not able to find an analytic methodology to adequately describe it, neither in the literature nor in our experience with the data set. 
Hence, we decided to tackle this problem through ML, and the first algorithm capable of automatically classifying the unstable modes in a physically meaningful groups (data clusters), that correspond to different plasma instability types, is given in this article. 

All of the described algorithms can be utilized separately, but are also combined in the Stability Analysis Vitalizing Instability Classification (\texttt{SAVIC}) code that can be used by anyone in the community for instability characterization in their own research. 
The code is user friendly and publicly available at the link provided in Section \ref{ssec:guide}. 
In this article, we only illustrate example applications that either provide important physical insights or highlight main features of \texttt{SAVIC}, with a complete description of the code is given in the documentation that accompanies the code release, alongside 28 figures. 
An overview description of \texttt{SAVIC} architecture and examples of its use are given in Section \ref{ssec:guide}. 
Finally, we use the results from our codes to illustrate the interplay between various types of unstable modes induced by either core protons, beam protons, or $\alpha$ particles, and their apparent hierarchical structure in governing the solar wind dynamics in Section \ref{sec:hierarchy}. 
The results given here provide us with all the required tools to build a comprehensive model of solar wind linear instabilities and their role in the solar wind evolution, which will be the topic of the next article in the series (Paper III).

%%%%%%%%%%%%%%%%%%%%%
%%% DATA & METHOD %%%
%%%%%%%%%%%%%%%%%%%%%

\section{Data and Methodology}
\label{sec:data}

Paper I describes the processing of the database provided by \cite{Durovcova_2019_SoPh}. 
Here, we briefly review the features of importance for this article. 
Approximately 1.5M ion VDFs are fitted as a sum of three generally anisotropic bi-Maxwellian (Equation \ref{eq:VDF}) VDFs---a proton core, proton beam, and $\alpha$ particles, with the beam and $\alpha$ populations having a drift with respect to the core 
\begin{equation}
    f_{j=c,b,\alpha} = \frac{n_j}{\pi^{3/2}w_{\perp,j}^2 w_{\parallel,j}} e^{-\frac{(v_\parallel - \Delta v,j)^2}{w_{\parallel,j}^2}} e^{-\frac{v_\perp^2}{w_{\perp,j}^2}}
    \label{eq:VDF}
\end{equation}
We label the thermal velocity as $w_{\perp,\parallel,j}=\sqrt{2 k_b T_{\perp,\parallel,j}/m_j}$ and the drift between the core and population $j$ as $\Delta v,j$. 
Here $k_b$ is the Boltzmann constant, and $m_j$, $n_j$, and $T_{\perp,\parallel}$ are the particle mass, density, and temperature for each VDF component respectively. 
Neither of the non-core populations necessarily needs to be identified in the fitting routine, dividing the data into four subsets: core only (C), core and beam (CB), core and $\alpha$ (C$\alpha$), and core, beam, and $\alpha$ (CB$\alpha$); see Table 1 in Paper I. 
The ion VDFs were sampled over a period of about one solar cycle (1974-1985) by the two \emph{Helios} spacecraft equipped with I1a and I1b particle analyzers \citep{Schwenn_1975}. 
In general, usage of linear dispersion solvers (see e.g. \cite{Roennmark_1982,Quataert_1998,Verscharen_2018_ArXiv_NHDS}) enables identifying a wave mode at a particular location in the wavevector / frequency space. 
The \texttt{PLUME} solver \citep{Klein_2015_PhPl} can be applied to the set of observed VDF parameters $\mathcal{P}$ to provide these solutions. 
The dimensionless parameters that comprise $\mathcal{P}$ include the core proton parallel plasma beta $\beta_{\parallel,j}= 2 \mu_0 n_j k_b T_{\parallel,j}/B^2$ where $\mu_0$ is magnetic permeability of vacuum and B is the magnetic field intensity, the temperature anisotropies of each component, the temperature disequilibrium between the components as well as their relative densities and drifts (see Equations 1 and 2 or Paper I).
In Paper I, we use its complement, \texttt{PLUMAGE} software, which performs contour integration of the dispersion relation $\underline{\mathbf{D}} \big( \omega, \mathbf{k} \big)$, where $\omega=\omega_{\textrm{r}} + i \gamma$, over the upper half of the complex frequency domain to determine if a given VDF is stable or unstable \citep{Klein_2017_JGRA}. 
The contour integration limits can be adjusted to increasingly large values of $\gamma$ to identify the Most Unstable Mode (MUM). 
The \texttt{PLUMAGE} code determines basic information about the MUM: growth rate normalized to proton gyrofrequency $\gamma^{\textrm{max}}/\Omega_p$, real frequency $\omega_{\textrm{r}}^{\textrm{max}}/\Omega_p$, wavevector normalized to gyroradius of core protons $\mathbf{k}^{\textrm{max}} \rho_c$, and the angle between wave propagation and the magnetic field $\theta_{kB}^{\textrm{max}}$; which is then fed back into \texttt{PLUME}, finding the detailed mode properties: e.g. electromagnetic eigenfunctions and estimated emitted power for each component, completing the set of the MUM wave parameters provided by \texttt{PLUME}, which we will refer to as \pplume. 
The core proton gyroradius and cyclotron frequency are given as $\rho_c = m_p w_{\perp,c} / e_c B$ and $\Omega_p = e_c B / m_p$, where $e_c$ is elementary charge. 
For each  VDF, the \texttt{PLUME} stability analysis provides between 11 and 21 output variables, depending on the number of fitted VDF components. 

Such high dimensionallity was the main motivation for introducing ML for stability analysis. 
Three types of algorithms are used in this work: 1) classification---determining if a given VDF is stable or unstable, and identifying the emitting component; 2) regression---evaluating \pplume for a given VDF, and; 3) clustering---characterizing different types of unstable modes within each subset. 

Both classification and regression were performed using the supervised Extreme Gradient Boosting (XGB) learning algorithm \citep{Chen_2015_XGB}. 
This powerful, constantly evolving open source code \citep{XGB}, is a scalable, parallel distributed gradient-boosted (GB) decision tree. 
In general, a decision tree creates a model that predicts the desired solution by evaluating a tree-like cascade of logical levels branched via \emph{if-then-else} True/False prompts, estimating the minimum number of questions needed to assess the probability of making a correct decision. 
GB algorithms create a number of different models and combine them into a single more accurate model based on the gradient of the error. 
The final prediction is a weighted sum of all of the separate tree predictions. 
The innovation introduced by XGB is that trees are built in parallel, instead of sequentially like in traditional GBs, scanning across gradient values for each new branched level. 
This way, XGB makes use of CPU/GPU resource parallelization to train the tree levels of the required accuracy within a reasonable amount of time, which would not be possible with other algorithms. 
This approach is probably the reason why for our data set, which has vastly different levels of coverage across the multi-dimensional phase space, GB vastly over-performs various linear and polynomial algorithms (Section \ref{sec:ML}). 
Such superior performance of XGB is well-established for financing models \citep{Horemuz_2018_XGB}, and has opened the path for numerous applications in that sector \citep{Li_2022_XGB_hybrid}. 

The clustering is done via unsupervised Gaussian Mixture (GM)---the expectation-maximization algorithm for fitting mixture-of-Gaussian models in an arbitrary number of dimensions \citep{Bishop_2006_ML,McNicholas_2016_mixture}. 
This model, widely applied in fields like psychology \citep{Shahin_2019_emotion} and finances \citep{Hodoshima_2019_stock}, increasingly finds utility in plasma physics \citep{Dupuis_2020_ApJ}. 
The ``proximity'' of $\mathcal{P}$ and \pplume parameters in phase space determines the distribution of solutions over a predetermined number of clusters. 
The number of clusters was determined empirically for each of the subsets: C (4), CB (8), C$\alpha$ (6), CB$\alpha$ (12). 
The physical motivations behind these clusters are discussed in Section \ref{ssec:classification}. 

%%%%%%%%%%
%%% ML %%%
%%%%%%%%%%

\section{Stability Analysis via Machine Learning Algorithms}
\label{sec:ML}

Before embarking on the description of the ML methods in our work, we note that the next two subsections are almost completely technical, with very limited physical insight; a reader interested only in physical interpretations may skip directly to Section \ref{ssec:classification}.

%%%%%%%%%%%%%%%%%%
%%% PREDICTION %%%
%%%%%%%%%%%%%%%%%%

\subsection{\texttt{SAVIC-P} - Predicting the plasma VDF stability}
\label{ssec:prediction}

\begin{figure*}
\centering
  \includegraphics[width=0.45\textwidth]{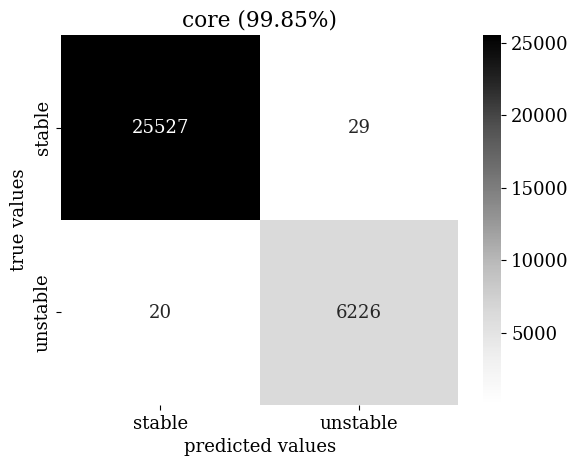}
  \includegraphics[width=0.45\textwidth]{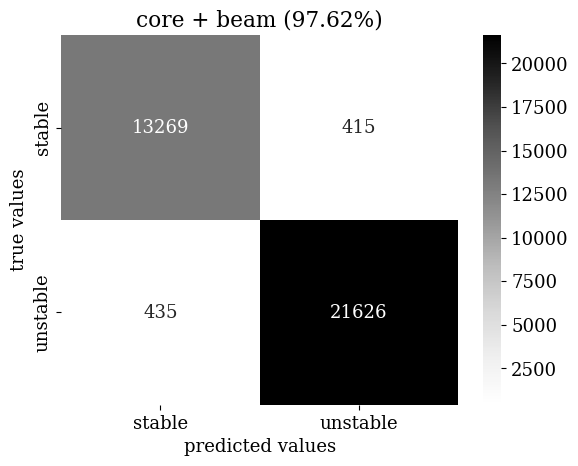}
  
  \includegraphics[width=0.45\textwidth]{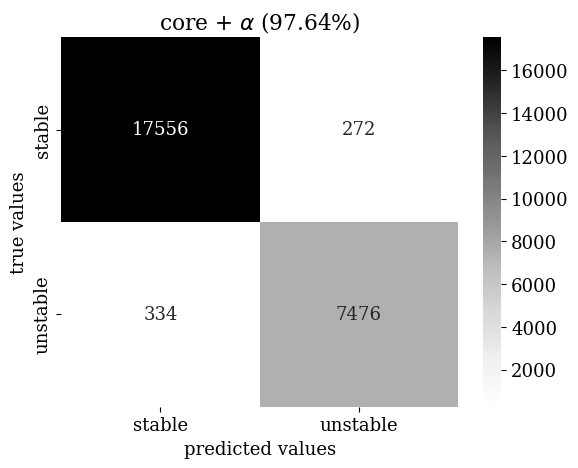}
  \includegraphics[width=0.45\textwidth]{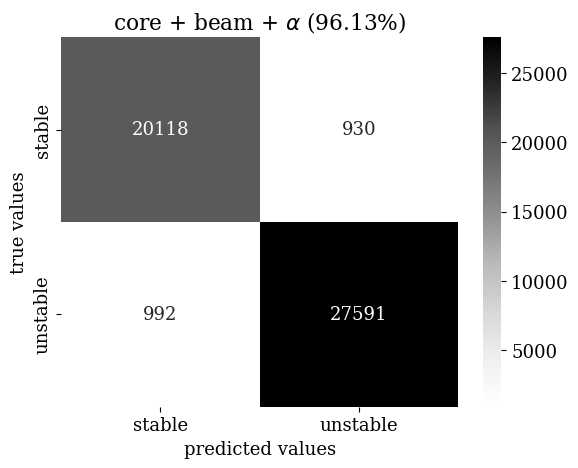}
  \caption{Confusion Matrices of plasma stability predicted by \texttt{SAVIC-P} for the four ion VDF subsets, compared against \texttt{PLUMAGE}-derived instability calculations as the "true" values. }
%    \caption{Confusion Matrices of plasma stability predicted by \texttt{SAVIC-P} for the four ion VDF subsets, where \pplume provides ``true'' values. }
  \label{fig:CM}
\end{figure*}

\begin{figure*}
\centering
  \includegraphics[width=0.95\textwidth]{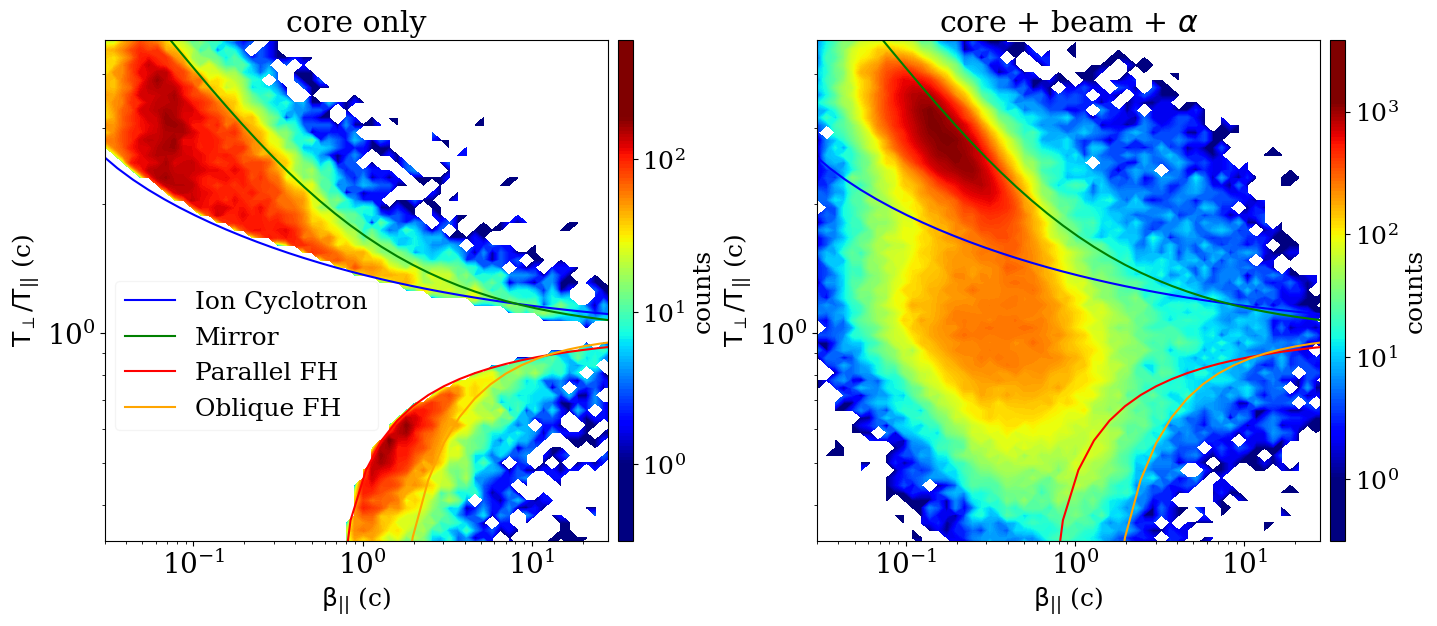}
  \caption{Comparison of the phase space densities of the unstable VDFs as a function of $\beta_{\parallel,c}$ and core temperature anisotropy for intervals when only core (\emph{left}) and where all three ion components (\emph{right panel}) are detected. 
  Solid lines show analytical thresholds for core anisotropy instabilities---Ion Cyclotron, Mirror, and Parallel and Oblique Firehose (FH) for $\gamma^{\textrm{max}}/\Omega_p = 10^{-4}$, following \cite{Verscharen_2016_ApJ}. 
  The parametric description of IC instability threshold (blue line) slightly differs from \texttt{PLUMAGE} predictions on \emph{left} panel, especially at low $\beta_\parallel,c$. 
  This difference causes the \texttt{SAVIC-P} to be more accurate if not aided by the parametric curves. 
  }
  \label{fig:prediction_c_cba}
\end{figure*}

For the set of parameters $\mathcal{P}$ associated with a given VDF, the first step is predicting if it is capable of generating any unstable modes. 
If it is not, than the VDF is classified as stable. 
We train four prediction algorithms associated with the four data subsets (C, CB, C$\alpha$, CB$\alpha$) using using 90\% of the available data, and perform testing using the remaining 10\%. 
Following the numbers provided in Table 1 of Paper I, the sizes of the four training sets are $\sim$54K, $\sim$195K, $\sim$67K, and $\sim$252K, respectively. 
The confusion matrices for the four subsets shown in Figure \ref{fig:CM}. 
The train / test ratio is arbitrary, and reducing the training set down to $\sim40\%$ of all data does not affect the accuracy of the predictions by more than a fraction of a percent. 

The prediction is notably more accurate for the case of a single proton population (one anisotropic Maxwellian) than for the other subsets. 
This feature is fairly easy to understand. 
Figure \ref{fig:prediction_c_cba} shows the unstable intervals on a traditional ``Brazil'' plot \citep{Kasper_2002_GRL}. 
The number of unstable modes that can arise for a single Maxwellian is limited, and their constraints are well described with analytical expressions for the temperature anisotropy as a function of plasma $\beta_\parallel$ (see, e.g. \cite{Verscharen_2016_ApJ}). 
These analytical expressions, shown in Figure~\ref{fig:prediction_c_cba}, are very accurate near moderate values of plasma $\beta_\parallel$ near unity where they have been historically applied, but are less accurate in lower- and higher-$\beta$ plasmas.
For this reason, these parametric curves were not used to aid the training algorithm. 
Including these expressions in testing versions of \texttt{SAVIC-P} reduced its accuracy, compared to the code described here. 
Introducing beam and $\alpha$ components drastically increases the number of potential free energy sources (see the list of $\mathcal{P}$ parameters in the \texttt{SAVIC-P} \emph{input} columns of Table \ref{tab:code_example}), and consequently the potential number of unstable modes to be encountered, scattering the stability margins over a large number of dimensions in the phase space. 
The relatively small population in some parts of this multi-dimensional space (e.g. low $\beta_\parallel,c$) is the main reason why \texttt{SAVIC-P} accuracy drops by a few percent.

%%%%%%%%%%%%%%%%%%%%%%
%%% QUANTIFICATION %%%
%%%%%%%%%%%%%%%%%%%%%%

\subsection{\texttt{SAVIC-Q} - Quantifying the Instability Parameters}
\label{ssec:quantifyication}

\begin{figure*}
\centering
  \includegraphics[width=0.9\textwidth]{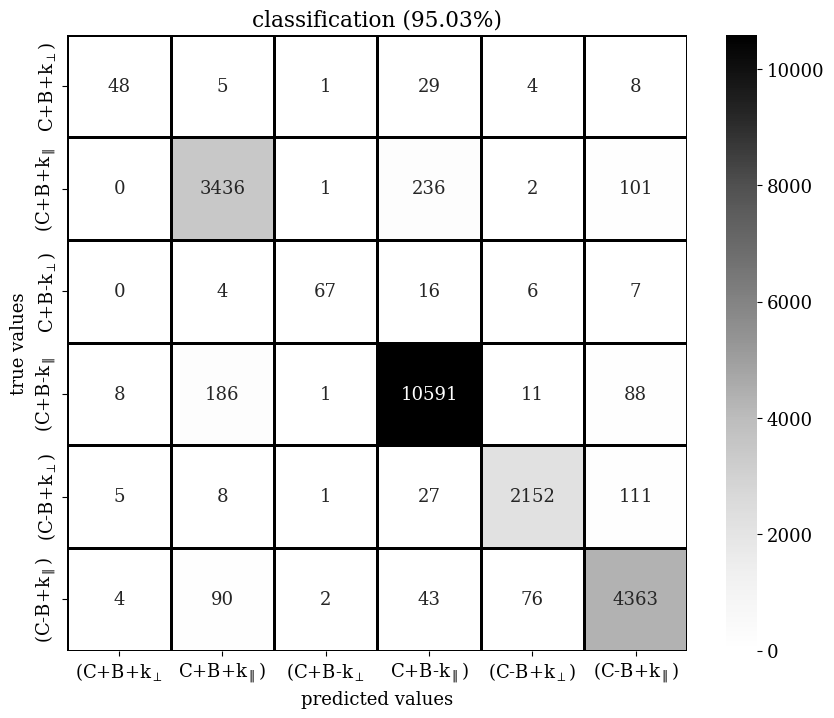}
  \caption{Classification of unstable modes for the CB data subset. 
  The "+" sign for core and beam signals that the component is emitting power, while $k_\perp$ and $k_\parallel$ stand for oblique and parallel propagation, respectively. 
  The groups that share the same \texttt{SAVIC-Q} regressor are marked by brackets. 
  }
  \label{fig:preclassify_cb}
\end{figure*}

\begin{figure*}
\centering
  \includegraphics[width=0.95\textwidth]{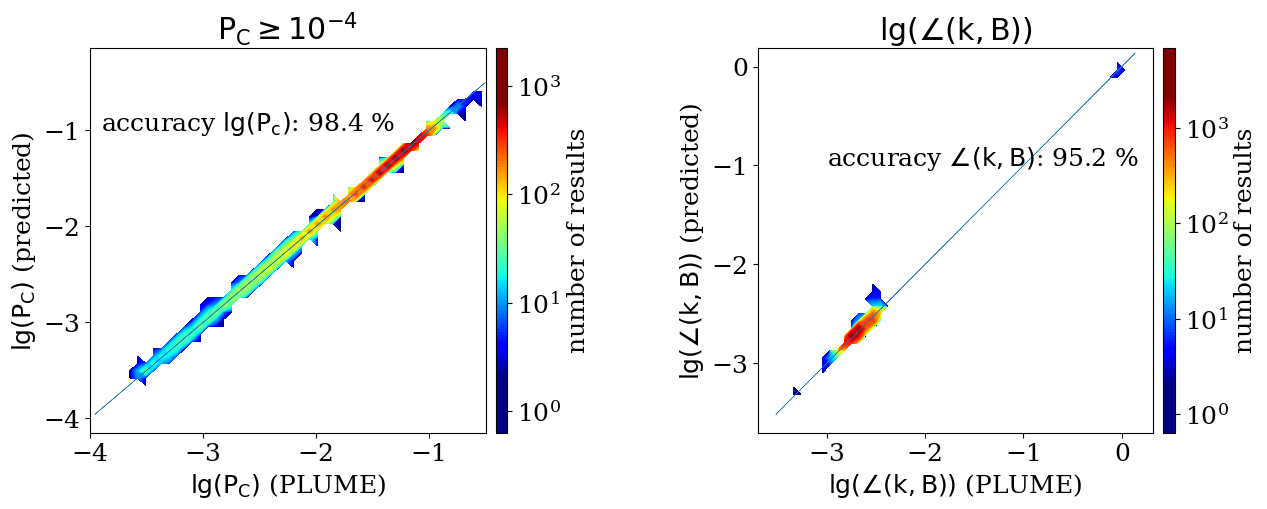}
  \includegraphics[width=0.95\textwidth]{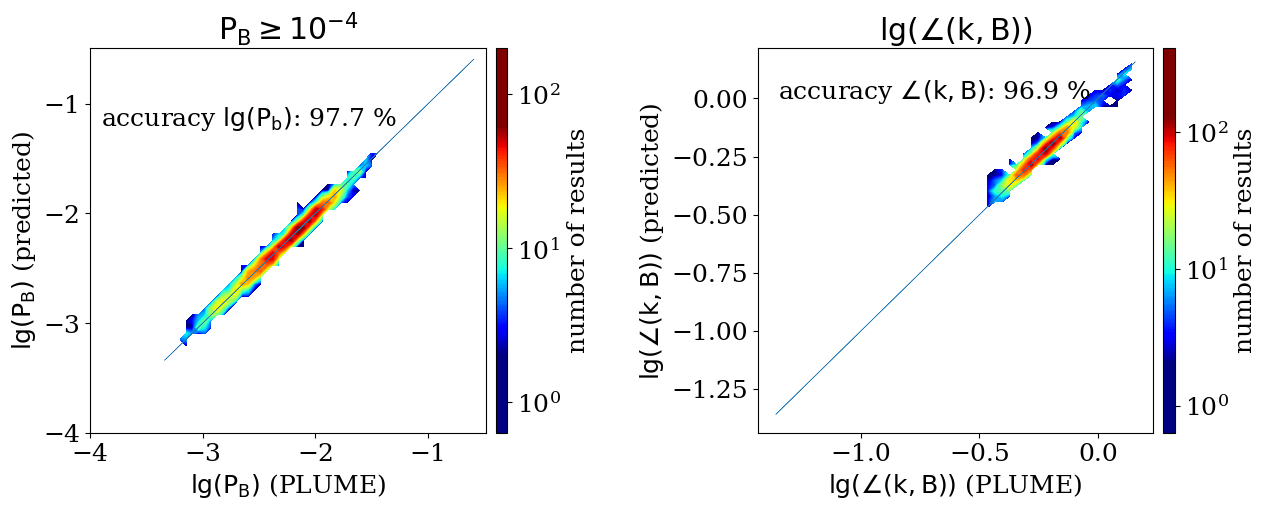}
  \caption{Examples of \texttt{SAVIC-Q} regression predictions for the CB data subset. 
  In the \emph{top row}, produced by $\mathrm{C+B-}$ regressor, we process the intervals where only the core is the emitting component, corresponding to data from \emph{third} and \emph{fourth} row on Figure \ref{fig:preclassify_cb}. 
  The \emph{bottom row} ($\mathrm{C-B+k_\perp}$ regressor) corresponds to \emph{fifth} row of Figure \ref{fig:preclassify_cb}, where $P_B$ and $\theta_{kB}^{\textrm{max}}$ (in radians) are estimated only for oblique modes. 
  }
  \label{fig:regression_cb}
\end{figure*}

Once a VDF is deemed unstable, we can quantify the features of the MUM. 
For training of the \texttt{SAVIC-Q} algorithm, we use $\mathcal{P}$ to predict a subset of \pplume, specifically the angle $\theta_{kB}^{\textrm{max}}$ and the normalized emitted power levels for each component, $P_{C,B,\alpha} \approx \gamma_{C,B,\alpha} / \omega_{\mathrm{r}}$ (c.f. \S 11.8 of \citep{Stix_1992}). 
These variables are chosen as they are used as input for the classifier described in Section \ref{ssec:classification}, but the \texttt{SAVIC-Q} regressors can be expanded to predict the rest of the \pplume variable set if needed. 

The regression process is, in general, less accurate than classification. 
In this case, we diagnosed the primary source of uncertainty to be the very large range of the emitted power values. 
When a given ion component is detected as part of the distribution, but does not participate in the unstable behavior, than the calculated emitted or absorbed power is not exactly zero, but a very small numerical value. 
Consequently, $P_{C,B,\alpha}$ varies by up to 10 orders of magnitude, and can be positive (representing emission) or negative (absorption). 
Traditionally, performance of a regressor decreases significantly if it is required to process such a large range of input values. 

To overcome this problem, we introduce another classifier within \texttt{SAVIC-Q}, prior to regression, that determines for a given unstable VDF and MUM, which components emit energy. 
An example for the CB subset is given in Figure \ref{fig:preclassify_cb}, again for a training sample containing 90\% of the subset. 
We separate the regressor algorithms into cases where the core and beam components are either emitting (+) or not emitting (-) power, with the wave propagation being either parallel ($k_\parallel$) or oblique ($k_\perp$) with respect to the magnetic field (\texttt{SAVIC-Q} \emph{output I} column of Table \ref{tab:code_example}). 
As the ``C-B-'' scenario is just a stable interval, up to six regressors can be trained. 
Two of the groups tabulated in the \emph{first} and \emph{third} rows (``C+B+$\mathrm{k_\perp}$'' and ``C+B-$\mathrm{k_\perp}$'') do not have sufficient data to train an accurate regressor, and are thus merged with their $k_\parallel$ counterparts. 
Once the intervals are grouped by the sign of the emitted power from each of the components, we can use the logarithm of $P_{C,B,\alpha}$ to decrease the span of the input. 
Following \citep{Klein_2019_ApJ} and Paper 1, we consider $P_C, P_B < 10^{-4}$ to be zero. 
Henceforth, we have built all of our models to use the logarithmic values of variables whenever possible---check the documentation for details. 

Results of the \texttt{SAVIC-Q} quantification are shown on Figure \ref{fig:regression_cb}, with specific examples given in the \emph{output II} columns of Table \ref{tab:code_example}.
The \emph{top panels} correspond to the \emph{third} and \emph{fourth} rows of Figure \ref{fig:preclassify_cb}---the proton beam is expected to either not participate in driving the MUM, or to absorb some of the emitted power. 
The span of the $\theta_{kB}^{\textrm{max}}$ angle is large as we need to process both parallel and oblique modes, given the relatively small sample size for the latter, but the method still provides a satisfying accuracy of over 95\%. 
The performance of the regressor is improved if we observe only oblique ($\mathrm{k}_\perp$) modes (\emph{bottom panel}), with $\theta_{kB}^{\textrm{max}}$ values being concentrated over the range of only $\sim15^\circ$. 
Better estimates of $P_C$ compared to minor components $P_B$ and  $P_\alpha$ are not surprising, as there are only two dominantly influential parameters---$\beta_{\parallel,c}$ and core temperature anisotropy---while $P_B$ has four major parameters: beam temperature anisotropy, drift, beam/core density and temperature ratio, none of which can be ignored by the regressors. 
As particles from each of the VDF populations will interact with the MUM  \citep{Verscharen_2019_LRSP}, our \pplume parameter predictions from similar regressors within \texttt{SAVIC-Q}, e.g. ``C+'', ``C+B-'', ``C+$\alpha$-'', and ``C+B-$\alpha$-'', are incompatible with each other.
In total, we train 17 regressors: 1, 4, 4, and 8 for C, CB, C$\alpha$, CB$\alpha$ data sets, respectively. 
The details on all the regressors are given in the public \texttt{SAVIC} documentation. 

%%%%%%%%%%%%%%%%%%%%%%
%%% CLASSIFICATION %%%
%%%%%%%%%%%%%%%%%%%%%%

\subsection{\texttt{SAVIC-C} - Classification of Unstable Modes}
\label{ssec:classification}

\begin{figure*}
\centering
  \includegraphics[width=0.38\textwidth]{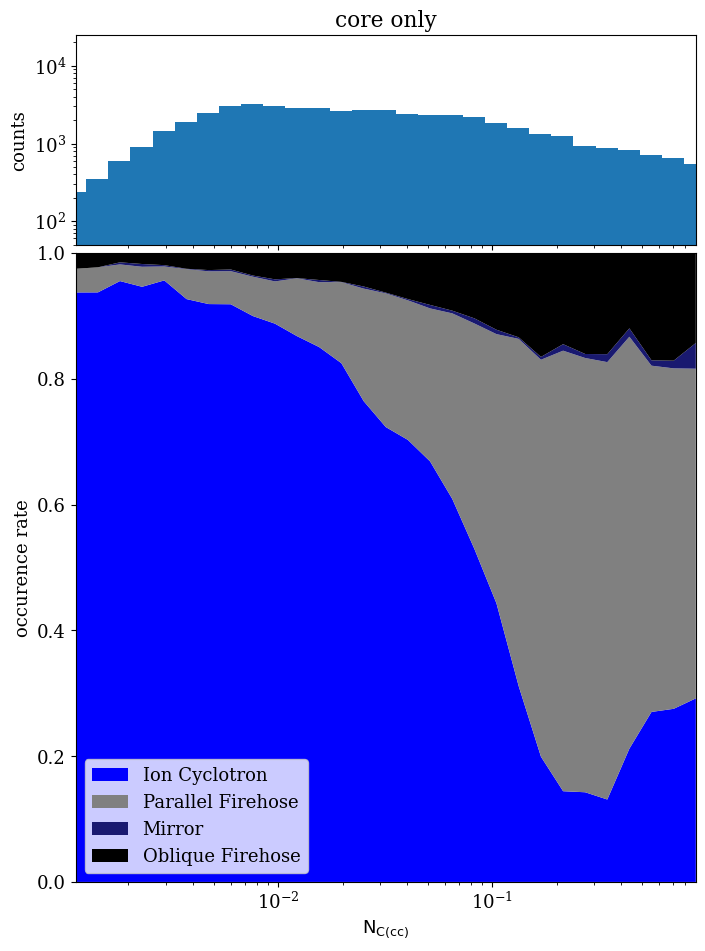}
  \includegraphics[width=0.38\textwidth]{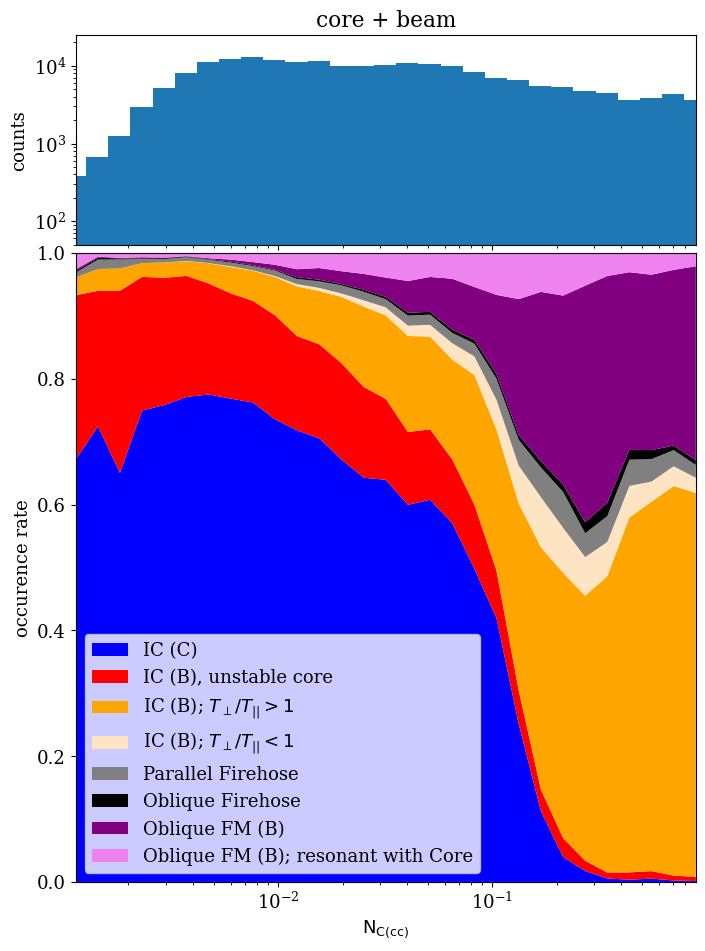}
  
  \includegraphics[width=0.38\textwidth]{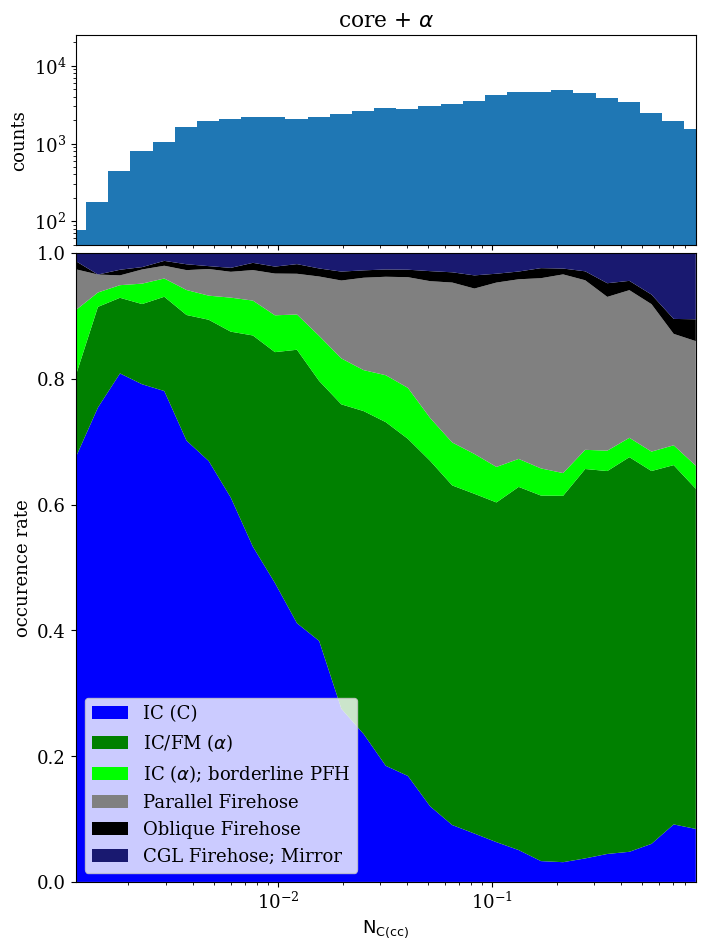}
  \includegraphics[width=0.38\textwidth]{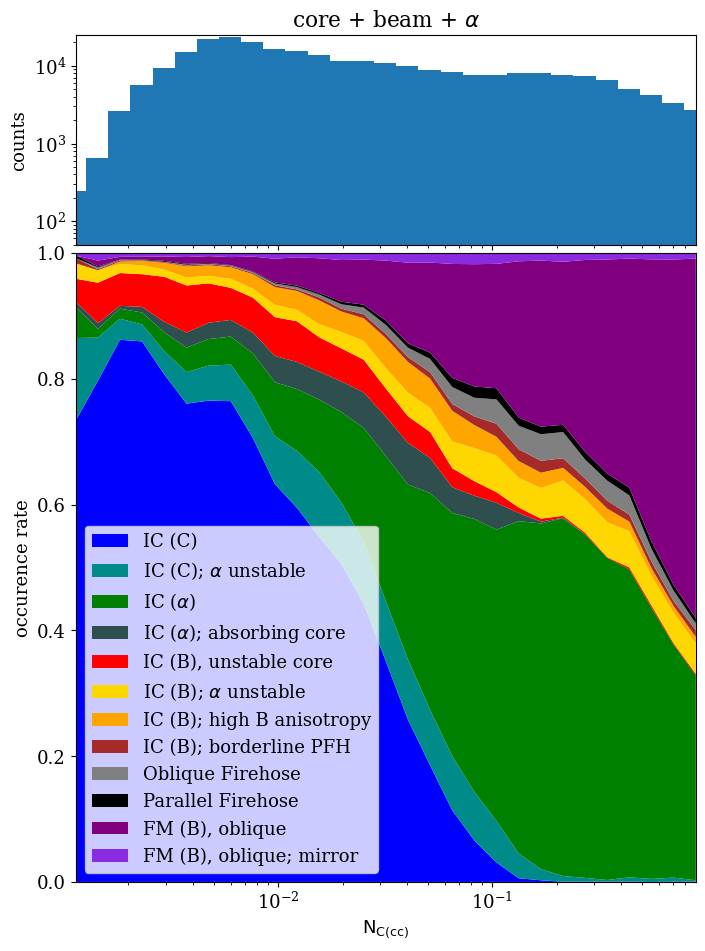}
  \caption{Overview of the instability clusters for all four subsets identified by \texttt{SAVIC-C}.
  The lack of $\alpha$ induced oblique modes is due to VDF fitting methods, which conflates core and beam populations of $\alpha$ particles into a single Maxwellian.
  Clarifications of different labels are given in Section \ref{ssec:classification}.
  }
  \label{fig:clusters}
\end{figure*}

The final, and physically most interesting, ML model aims to classify the modes predicted to be unstabled by \texttt{SAVIC-P} and with parameters quantified by \texttt{SAVIC-Q}, into groups through clustering.
Each cluster should ideally represent all the intervals where a certain type of instability (e.g. Ion Cyclotron (IC) or mirror mode) is active, also noting which component is emitting energy (e.g. (B) for beam on Figure \ref{fig:clusters}), and none of the intervals that do not feature this particular instability.  
Example outputs from \texttt{SAVIC-C} are found in the final column of Table \ref{tab:code_example} and as a function of $N_{C(cc)}$ in Figure \ref{fig:clusters}. 
Providing a label that describes the governing physical mechanisms, in addition to the numerical description of a predicted unstable mode, is a novel element of this code compared to traditional dispersion solvers. 

For each of the four subsets, the number of clusters used in the GM algorithm that underlies \texttt{SAVIC-C} is determined empirically. 
To find the appropriate number of clusters for each subset, we manually tested dozens of combinations of variables drawn from $\mathcal{P}$ and \pplume, settling on slightly different input sets for all four data subsets, with the entire $\mathcal{P}$ set, $P_C$, $\theta_{kB}^{\textrm{max}}$, and analytical thresholds for core instabilities---IC, mirror, Parallel (PFH) and Oblique Firehose (OFH)---\citep{Verscharen_2016_ApJ} used for all four data subsets. 
For the CB, C$\alpha$, and CB$\alpha$ subsets, we add $P_{B,\alpha}$ (when positive), and the analytical thresholds for the firehose instability for a multi-component plasma \citep{Chen_2016_ApJ}. 
It is important to note specific behavior of the models in relation to mode properties. 
Namely, if some of the \pplume variables are very similar for different instabilities due to their similar physical features, such as compressibility $\delta n$ or  $\gamma^{\textrm{max}}/\Omega_p$, than including these makes the clustering algorithm less accurate. 
There are several major technical features of the clustering results shown on Figure \ref{fig:clusters} that are worth emphasizing, while exhaustive discussion of physical implications will be given in Paper III. 

Core results are fairly straightforward to obtain, as there are only four major instabilities to consider. 
An interesting result is that the IC is the dominant instability for all the anisotropy values above unity, except for $\beta_{\parallel,c} \gtrsim 10$. 
The non-propagating mirror mode is, once triggered, expected to grow with anisotropy in a notably more rapid fashion than the IC for most values of $\beta_{\parallel,c}$ \citep{Gary_1993,Hellinger_2007_PhPl,Klein_2013_PhD} and to be the MUM for most intervals above the green line on the \emph{left panel} of Figure \ref{fig:prediction_c_cba}. 
On the contrary, it turns out that for most of the VDFs measured in the solar wind that are sensitive to mirror mode, the IC remains the MUM, as the realistic VDFs are both very far from the IC threshold (blue line) and sufficiently close to the mirror threshold (green line). 
We confirmed by visual inspection that \texttt{SAVIC-C} correctly classifies these modes. 

The clusters in the CB subset labeled as IC (B)---the beam is the most intensive emitting component causing the parallel propagating MUM---also contain an admixture of parallel fast modes (FMs). 
In almost all the intervals from these clusters, the IC mode is triggered by beam anisotropy (IC (B) $T_\perp / T_\parallel <> 1$) (either larger or lower than unity), and FM by the beam drift. 
In the collisionally young wind, IC dominates, and FM will not be detected as the MUM if there is beam anisotropy induced parallel IC mode present, unless the beam drift values are very high \citep{Daughton_1998_JGR}. 
In the collisionally old wind, when beam anisotropy is low enough to stop being a formidable source of free energy, and the drift is not strong enough to power the parallel FM, the marginally unstable distributions dominantly feature the slow growing oblique FM as MUM. 
It is important to emphasize that recognizing beams as separate, drifted component instead of working with the moments of the entire proton VDF is crucial for accurately predicting these modes \citep{Klein_2021_ApJ} and finding the agreement with in situ observations of local electric and magnetic field \citep{Vech_2021_AA}. 

The C$\alpha$ data set is comprised of 6 clusters instead of 8, primarily because $\alpha$ distribution is fitted as a single Maxwellian due to instrument range and resolution limits. 
Consequently, the fitting methods developed by \cite{Durovcova_2019_SoPh}---and in complementary work by \cite{Stansby_2018_SolPh}--- identify very large parallel temperatures $T_{\parallel,\alpha}$ in the young wind, as they are unable to separate drifted $\alpha$ beams. 
Unlike for the case of proton beams, about a third of the parallel modes given in \emph{green} on \emph{bottom left panel} of Figure \ref{fig:clusters} are FMs induced by the excess parallel pressure of the $\alpha$ component. 
It is also worth noting that \emph{light green} and \emph{dark blue} clusters in the \emph{bottom left panel} of Figure \ref{fig:clusters} are fundamentally different in nature. 
As discussed in detail in Paper I, the beam can sometimes be mis-identified as part of the core due to instrument limitations. 
This will lead to artificial increase in $T_{\parallel,c}$, which our clustering algorithm recognizes as Chew-Goldberger-Low (CGL) FH---a MHD instability caused by very strong pressure anisotropy \citep{Chew_1956_RSPSA}. 
On the other hand, the \emph{light green} cluster features the combination of two $\mathcal{P}$ components---$T_{\perp(\alpha)} / T_{\parallel,\alpha} \gtrsim 1$ and $T_{\perp(c)} / T_{\parallel,c} \lesssim 1$ very close to PFH threshold---where the core protons are just barely unable to create the FH instability, but are anisotropic enough to resonate with mildly drifted $\alpha$ component and absorb a fraction of the power emitted from the $\alpha$ population. 
This phenomena is characteristic for older wind, and its beam-core interaction analog is observed for CB$\alpha$, but this time with different type of phase space resonance with strongly drifted and highly anisotropic beams. 

The presence of the mild "background", oblique beam FM is present in the collisionally old wind for both the CB and CB$\alpha$ subsets. 
This mode can be ``resonant'' with the core in some cases---having the core absorb part of the energy emitted by the beam (see Section 6 of \cite{Verscharen_2019_LRSP}). 
Also, as it is mostly sampled in the young wind where the instrument performance is most optimal, CBA subset features some VDFs with highly anisotropic core component that have the mirror as MUM. 
This phenomena will be discussed in detail in the follow-up paper, where we will argue that this mode is the ever-present regulator of the beam drift in the solar wind. 
Due to large number of clusters within the CB$\alpha$ subset, and some of the clusters having very low number of intervals, a complete separation of modes was not achieved in all cases. 
For example, the cluster labeled as ``IC (B); borderline PFH'' contains 0.96\% of all the subset intervals, and contains two groups: weakly unstable ($\gamma^{\textrm{max}}/\Omega_p < 0.5 \cdot 10^{-3}$) IC (B) intervals and very weakly ``borderline'' PFH unstable intervals. 
Ideally, both of these groups should belong to their respective clusters, but this small cluster was still maintained within \texttt{SAVIC-C} as a remnant of the uncertainty of our method. 

\subsection{Public Stability Analysis Code Architecture and Usage Example}
\label{ssec:guide}

\begin{figure*}
\centering
    \begin{forest}
      for tree={
        grow'=0,
        draw, rounded corners,
        /tikz/align=center, 
        calign=center,
        anchor=center,
        font=\sffamily\bfseries\linespread{0.9}\selectfont,
        text=white,
        inner ysep=2pt,
        parent anchor=children,
        child anchor=parent,
        l sep=2mm,
        s sep=2mm,
        where level=0{text width=3em}{},
        where level=1{text width=4em}{},
        where level=2{text width=4em}{},
        where level=3{text width=4em}{},
        where level=4{text width=4em}{},
        where level=5{text width=6em}{},
        where level=6{text width=6em}{},
        where level=7{text width=5em}{},
        where level=8{text width=4em}{},
        if n children=0{
          tier=terminal
        }{}
      },
      clr/.style={fill=#1}
    [read in and \\ classify $\mathcal{P}$ ,clr=orange
        [core,clr=yellow!60!orange
            [\texttt{SAVIC-P}\\Stable / Unstable\\Classifier,clr=teal
                [VDF is stable,clr=gray
                    [report \& out,clr=black,text width=3em]
                ]
                [VDF is not stable,clr=gray, name=CVDFUS
                    [ ,phantom
                        [, phantom
                            [{\texttt{SAVIC-Q} (C) \\ $P_c$, $\angle (\mathbf{k},\mathbf{B})$ \\Regressor},clr=yellow!60!blue, name=regC, 
                                [\texttt{SAVIC-C} (C) GM Clustering,clr=green!60!blue,name=CGMC,
                                    [report \& out,clr=black,text width=3em]
                                ]
                            ]
                        ]
                    ]
                ]
            ]
        ]
        [core + beam,clr=yellow!60!orange
            [\texttt{SAVIC-P}\\Stable / Unstable\\Classifier,clr=teal
                [VDF is stable,clr=gray
                    [report \& out,clr=black,text width=3em]
                ]
                [VDF is not stable,clr=gray
                    [{\texttt{SAVIC-Q} $P_c$, $P_b$, $\angle (\mathbf{k},\mathbf{B})$ Classifier},clr=yellow!60!magenta
                        [{$P_c > 0$, $P_b > 0$, $\angle (\mathbf{k},\mathbf{B}) \to 0$},clr=yellow!60!green, name=clsC1B1k0,
                            [{\texttt{SAVIC-C} \\ $P_c$, $P_b$, $\angle (\mathbf{k},\mathbf{B})$ \\Regressor},clr=yellow!60!blue,phantom, 
                                [\texttt{SAVIC-C} (CB) GM Clustering,clr=green!60!blue,phantom]
                            ]
                        ]
                        [{$P_c > 0$, $P_b > 0$, $\angle (\mathbf{k},\mathbf{B}) \to 0$},clr=yellow!60!green,phantom,
                            [{\texttt{SAVIC-Q} (CB) \\ $P_c$, $P_b$, $\angle (\mathbf{k},\mathbf{B})$ \\Regressor},clr=yellow!60!blue, name=regC1B1, 
                                [\texttt{SAVIC-C} (CB) GM Clustering,clr=green!60!blue,phantom]
                            ]
                        ]
                        [{$P_c > 0$, $P_b > 0$, $\angle (\mathbf{k},\mathbf{B}) > 0$},clr=yellow!60!green, name=clsC1B1k1,
                            [{\texttt{SAVIC-C} \\ $P_c$, $P_b$, $\angle (\mathbf{k},\mathbf{B})$ \\Regressor},clr=yellow!60!blue,phantom, 
                                [\texttt{SAVIC-C} (CB) GM Clustering,clr=green!60!blue,phantom]
                            ]
                        ]
                        [{$P_c > 0$, $P_b < 0$, $\angle (\mathbf{k},\mathbf{B}) \to 0$},clr=yellow!60!green, name=clsC1B0k0, 
                            [{\texttt{SAVIC-C} \\ $P_c$, $\angle (\mathbf{k},\mathbf{B})$ \\Regressor},clr=yellow!60!blue, phantom, 
                                [\texttt{SAVIC-C} (CB) GM Clustering,clr=green!60!blue,phantom]
                            ]
                        ]
                        [{$P_c > 0$, $P_b < 0$, $\angle (\mathbf{k},\mathbf{B}) \to 0$},clr=yellow!60!green, phantom,
                            [{\texttt{SAVIC-Q} (CB) \\ $P_c$, $\angle (\mathbf{k},\mathbf{B})$ \\Regressor},clr=yellow!60!blue, name=regC1B0, 
                                [\texttt{SAVIC-C} (CB) GM Clustering,clr=green!60!blue,name=CBGMC,
                                    [report \& out,clr=black,text width=3em]
                                ]
                            ]
                        ]
                        [{$P_c > 0$, $P_b < 0$, $\angle (\mathbf{k},\mathbf{B}) > 0$},clr=yellow!60!green, name=clsC1B0k1,
                            [{\texttt{SAVIC-C} \\ $P_c$, $\angle (\mathbf{k},\mathbf{B})$ \\Regressor},clr=yellow!60!blue,phantom, 
                                [\texttt{SAVIC-C} (CB) GM Clustering,clr=green!60!blue,phantom]
                            ]
                        ]
                        [{$P_c < 0$, $P_b > 0$, $\angle (\mathbf{k},\mathbf{B}) \to 0$},clr=yellow!60!green,
                            [{\texttt{SAVIC-Q} (CB) (1) \\ $P_b$, $\angle (\mathbf{k},\mathbf{B})$ \\Regressor},clr=yellow!60!blue, name=regC0B1k0, 
                                [\texttt{SAVIC-C} (CB) GM Clustering,clr=green!60!blue,phantom]
                            ]
                        ]
                        [{$P_c < 0$, $P_b > 0$, $\angle (\mathbf{k},\mathbf{B}) > 0$},clr=yellow!60!green,
                            [{\texttt{SAVIC-Q} (CB) (2) \\ $P_b$, $\angle (\mathbf{k},\mathbf{B})$ \\Regressor},clr=yellow!60!blue, name=regC0B1k1, 
                                [\texttt{SAVIC-C} (CB) GM Clustering,clr=green!60!blue,phantom]
                            ]
                        ]
                    ]
                ]
            ]
        ]
        [core + $\alpha$,clr=yellow!60!orange
            [\texttt{SAVIC-P}\\Stable / Unstable\\Classifier,clr=teal
                [VDF is stable,clr=gray
                    [report \& out,clr=black,text width=3em]
                ]
                [VDF is not stable,clr=gray, name=CAVDFUS
                    [ ,phantom
                        [, phantom
                            [\texttt{SAVIC-Q}  (C$\alpha$) \\ Logarithmic Regressors, name=intCA, clr=red 
                                [\texttt{SAVIC-C} (C$\alpha$) GM Clustering,clr=green!60!blue,name=CAGMC,
                                    [report \& out,clr=black,text width=3em]
                                ]
                            ]
                        ]
                    ]
                ]
            ]
        ]
        [core + beam + $\alpha$,clr=yellow!60!orange
            [\texttt{SAVIC-P}\\Stable / Unstable\\Classifier,clr=teal
                [VDF is stable,clr=gray
                    [report \& out,clr=black,text width=3em]
                ]
                [VDF is not stable,clr=gray, name=CBAVDFUS
                    [ ,phantom
                        [, phantom
                            [\texttt{SAVIC-Q}  (CB$\alpha$) \\ Logarithmic Regressors, name=intCBA, clr=red 
                                [\texttt{SAVIC-C} (CB$\alpha$) GM Clustering,clr=green!60!blue,name=CBAGMC,
                                    [report \& out,clr=black,text width=3em]
                                ]
                            ]
                        ]
                    ]
                ]
            ]
        ]
    ]
    \draw (clsC1B1k1) -- (regC1B1);
    \draw (clsC1B1k0) -- (regC1B1);
    \draw (clsC1B0k1) -- (regC1B0);
    \draw (clsC1B0k0) -- (regC1B0);
    \draw (regC1B1) -- (CBGMC);
    \draw (regC0B1k1) -- (CBGMC);
    \draw (regC0B1k0) -- (CBGMC);
    \draw (CVDFUS) -- (regC);
    \draw (CAVDFUS) -- (intCA);
    \draw (CBAVDFUS) -- (intCBA);
    \end{forest}
  \caption{The \texttt{SAVIC} algorithm of plasma instabilities prediction, quantification, and classification. 
  C$\alpha$ and CB$\alpha$ classifiers and regressors are suppressed into single boxes for simplicity. }
  \label{fig:algorithm}
\end{figure*}
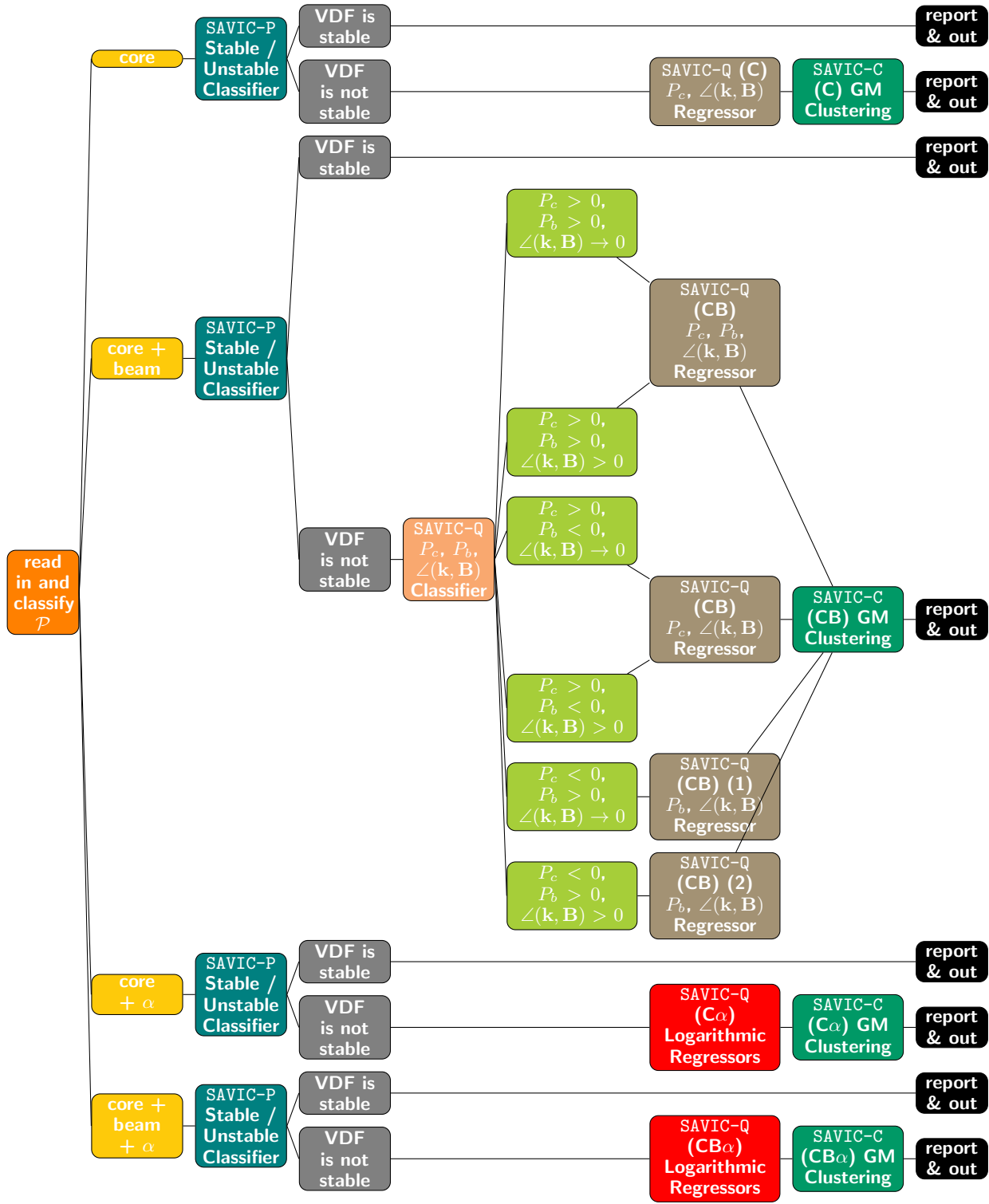

The three parts of the \texttt{SAVIC} code---stability predictor (\texttt{SAVIC-P}), quantifying classifier / regressor (\texttt{SAVIC-Q}), and unstable mode classifier (\texttt{SAVIC-C})---presented in Sections \ref{ssec:prediction}-\ref{ssec:classification} are available at  \url{https://github.com/MihailoMartinovic/SAVIC}. 
They can be used separately, but are also incorporated in a chain that provides a full analysis of a given VDF. 
Here, we will first present the concept of its use following the CB scheme in Figure \ref{fig:algorithm}, and then provide an illustrative example. 

The input contains the information about one or more VDFs to be processed. 
The format is explained in Paper I, the code documentation, and is also given in the example below. 
The VDF parameters are read and categorized into one of the four subsets. 
The \texttt{SAVIC-P} classifier described in Section \ref{ssec:prediction} then determines if the distribution is stable or unstable (\emph{third column}). 
If it is stable, the algorithm ends. 
Otherwise, a second classifier within \texttt{SAVIC-Q} described in Section \ref{ssec:quantifyication} is engaged, determining if the mode is parallel or oblique, and if the emitting component is core, beam, or both (\emph{light orange, fourth column}). 
Based on that information, the data is sent into one of four (for the CB case) \texttt{SAVIC-C} logical regressors, given in \emph{light green} in \emph{fifth column}. 
The output of this step is the emitted power, and the wave vector propagation angle, which is, along with $\mathcal{P}$, enough information to feed the \texttt{SAVIC-C} classifier described in Section \ref{ssec:classification} (\emph{sixth column, green}). 
The final output includes the information on emitting components, direction of propagation, and a type of unstable mode.

%\begin{table*}
    %\centering
    %\begin{tabular}{|cccccc|c|c|ccc|c|}
    %\hline
        %\multicolumn{7}{|c}{\texttt{SAVIC-P}} & \multicolumn{4}{|c|}{\texttt{SAVIC-Q}} & SAVIC-C \\
        %\hline
        %$\beta_{\parallel,c}$ & $\frac{T_{\perp,c}}{T_{\parallel,c}}$ & $\frac{T_{\parallel,c}}{T_{\parallel,b}}$ & $\frac{T_{\perp,b}}%{T_{\parallel,b}}$ & $\frac{n_b}{n_c}$ & $\frac{\Delta v_{b,c}}{v_{Ac}}$ & unstable & mode class &  $P_C$ &  $P_B$ &  %$\theta_{kB}^{\textrm{max}}$ &  mode type \\
        %\hline
        %1.0 &    1.0 &   1.0 &    1.0 &  0.05 &  0.5 &     False &      -- &    -- &        -- &      -- &                           --  \\
        %1.5 &    \emph{2.5} &   0.8 &    1.0 &  0.05 &  0.5 &      True &  C+B-k- &  0.19 &        -- &  0.0041 &         IC (C) \\
        %0.5 &    1.0 &   1.0 &    \emph{3.5} &   0.1 &  1.5 &      True &  C-B+k- &    -- &      0.12 &  0.0039 &  IC (B); $T_\perp/T_{||} > 1$ \\
        %0.8 &    1.1 &   1.0 &    1.2 &  0.05 &  \emph{2.1} &      True &  C-B+k+ &    -- &     0.005 &  0.8451  &  FM (B), oblique \\
        %0.5 &    0.7 &   0.8 &    0.8 &  0.01 &  0.2 &     False &      -- &    -- &        -- &      -- &                          --   \\
        %0.8 &    \emph{3.1} &   1.0 &    \emph{3.9} &   0.1 &  1.9 &      True &  C+B+k- &  0.21 &    0.0007 &   0.0010 &         IC (B), unstable %core \\
%        \hline
%    \end{tabular}
%    \caption{Example of the \texttt{SAVIC} code usage for CB VDFs. }
%    \label{tab:code_example1}
%\end{table*}

\begin{deluxetable*}{|cccccc|c|c|ccc|c|}
\label{tab:code_example}
\tablecaption{Example of the \texttt{SAVIC} code usage for VDFs from the CB subset.}
\tablehead{
\multicolumn{6}{|c|}{} & \texttt{SAVIC-P} &  \texttt{SAVIC-Q} & \multicolumn{3}{|c|}{ \texttt{SAVIC-Q} \hspace{1mm} \texttt{SAVIC-C}} & \texttt{SAVIC-C} \\
\multicolumn{6}{|c|}{$\mathcal{P}$---input I for all codes} & output & output I \& input II & \multicolumn{3}{|c|}{output II \hspace{1mm} input II} & output
}
\startdata
        $\beta_{\parallel,c}$ & $\frac{T_{\perp,c}}{T_{\parallel,c}}$ & $\frac{T_{\parallel,c}}{T_{\parallel,b}}$ & $\frac{T_{\perp,b}}{T_{\parallel,b}}$ & $\frac{n_b}{n_c}$ & $\frac{\Delta v_{b,c}}{v_{Ac}}$ & unstable & mode class &  $P_C$ &  $P_B$ &  $\theta_{kB}^{\textrm{max}}$ &  MUM type \\
        \hline
        1.0 &    1.0 &   1.0 &    1.0 &  0.05 &  0.5 &     False &      -- &    -- &        -- &      -- &                           --  \\
        1.5 &    \emph{2.5} &   0.8 &    1.0 &  0.05 &  0.5 &      True &  C+B-k$_\parallel$ &  0.19 &        -- &  0.0041 &         IC (C) \\
        0.5 &    1.0 &   1.0 &    \emph{3.5} &   0.1 &  1.5 &      True &  C-B+k$_\parallel$ &    -- &      0.12 &  0.0039 &  IC (B); $T_\perp/T_{||} > 1$ \\
        0.5 &    0.5 &   \emph{2.9} &    2.4 &  0.08 &  \emph{1.5} &      True &  C-B+k$_\perp$ &    -- &     0.012 &  0.57  &  FM (B), oblique \\
        0.5 &    0.7 &   0.8 &    0.8 &  0.01 &  0.2 &     False &      -- &    -- &        -- &      -- &                          --   \\
        0.8 &    \emph{3.1} &   1.0 &    \emph{3.9} &   0.1 &  1.9 &      True &  C+B+k$_\parallel$ &  0.21 &    0.0007 &   0.0010 &         IC (B), unstable core \\
        \hline
\enddata
\end{deluxetable*}

An example can be given as follows. 
The input data is in the first section of Table \ref{tab:code_example}. 
The \texttt{SAVIC-P} finds that four out of six VDFs are unstable, while \texttt{SAVIC-Q} diagnoses which component emits energy. 
Each MUM has its main free energy source---a population of particles within either core or beam that is sufficiently far away from LTE to intensively emit energy in situ. 
In this case, the energy is emitted due to either strong core/beam anisotropy, or the beam moving much faster than the core. 
The VDF parameter responsible for the instability is \emph{italicized} in the table. 
For each unstable interval, the appropriate  \texttt{SAVIC-Q} regressor (Section \ref{ssec:quantifyication}) is launched to find the power from the emitting component(s), and the propagation angle. 
Finally, all the obtained information is fed into the GM clustering algorithm, which provides the mode description (Section \ref{ssec:classification}). 

All the described steps are automatized, and therefore activated by a single user command for an arbitrarily large data set. 
The  \texttt{SAVIC} code is extremely efficient as it uses already trained ML entities, and can process millions of VDFs in only seconds of real time.

%%%%%%%%%%%%%%%%%
%%% HIERARCHY %%%
%%%%%%%%%%%%%%%%%

\section{Hierarchical Structure of Solar Wind Instabilities}
\label{sec:hierarchy}

\begin{figure*}
\centering
  \includegraphics[width=0.6\textwidth]{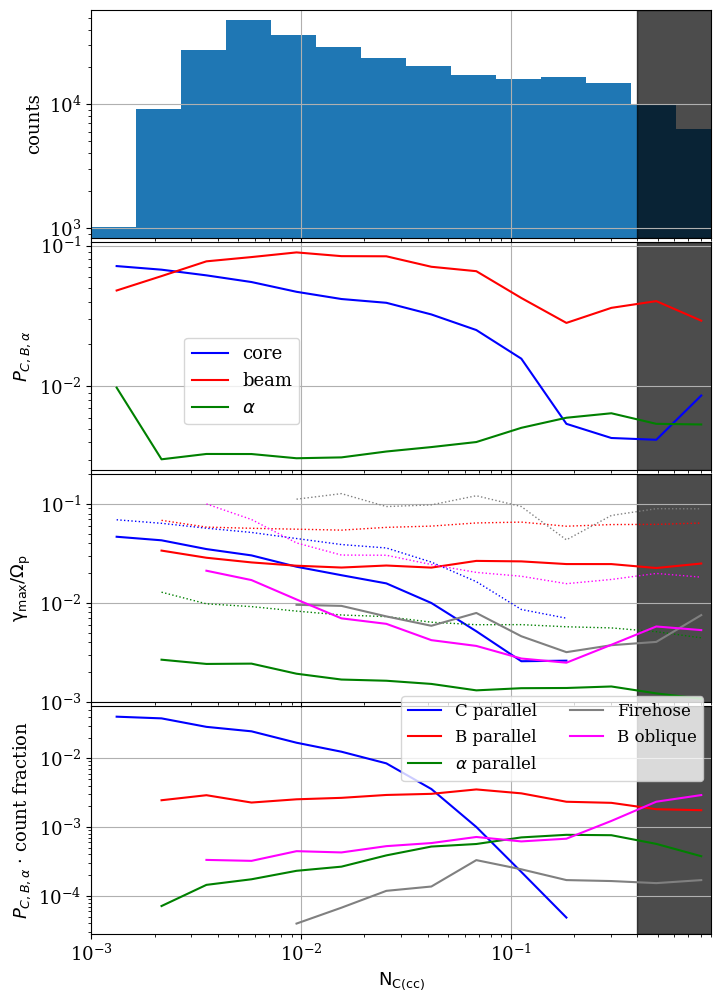}
  \caption{Quantification of the contribution of the five groups of unstable modes to the solar wind stability dynamics: parallel modes caused by core, beam, or $\alpha$ particles, beam oblique modes, and FH instabilities. 
  \emph{Second panel} shows medians of positive values of $P_C$, $P_B$, and $P_\alpha$. 
  In the \emph{third panel}, medians and 80th percentiles of the MUM growth rate are given in solid and dotted lines, respectively. 
  Finally, the total contribution of each group is quantified in the \emph{bottom panel}, where the emitted power is normalized by the relative number of occurrences for each $N_{C(cc)}$ bin (for the data point to be shown, there must be at least 100 intervals from a group in a given bin). 
  The part of phase space where \emph{Helios} instruments have limited reliability is shaded in \emph{grey}.
  }
  \label{fig:power_nc_cba}
\end{figure*}

Each of the algorithms presented in Section \ref{sec:ML} can be used not just for the overarching description of the solar wind linear instabilities, as we aim to do in Paper III, but also for addressing any stability related project that would otherwise require millions of CPU hours consumed by a powerful dispersion solver, such as \texttt{PLUME}, to process any statistically large data set. 
For example, even though \texttt{PLUME} and \texttt{PLUMAGE} solvers are highly optimized, producing the training data set used here and described in Paper I required $\sim$8M CPU hours. 
In this Section, we demonstrate the utility of the clustering algorithm (Section \ref{ssec:classification}) by investigating the overall interplay between different types of instabilities as the solar wind is being gradually processed by collisions. 

The CB$\alpha$ data set has 12 identified mode types (Figure \ref{fig:clusters}, \emph{bottom right}). 
To illustrate the evolution of the these modes, we merge them into 5 groups: parallel modes driven by any of the three components, beam induced oblique modes, and FH modes. 
We group them this way to illustrate the potential of using \texttt{SAVIC-C} by a user with developed physical intuition regarding a given problem. 
Moreover, addressing 12 separate modes in detail is beyond the scope of this paper, and would be a tedious process with little additional physical insight for this particular example. 
On the \emph{second panel} of Figure \ref{fig:power_nc_cba} we show median values of $P_C$,  $P_B$, and  $P_\alpha$. 
It is important to note that if a MUM is induced by a single component (e.g. beam), \pplume also contains information about the power from other two components that might be positive or negative, e.g. \emph{red} and \emph{violet} areas on \emph{upper right panel} of Figure \ref{fig:clusters}, respectively. 
For simplified analysis, we only take the positive power values into account in the \emph{second panel} of Figure \ref{fig:power_nc_cba}. 
We also mark the collisionally old sector of the wind where \emph{Helios} observations have limited confidence, which is addressed in detail in Paper I. 

The median of $P_B$ is the highest of the three everywhere except in very young solar wind, while $P_\alpha$ is almost constantly the lowest. 
The apparent increase in $P_\alpha$ in older wind is the VDF fitting effect explained in Section \ref{ssec:classification}. 
This simple approach would suggest that the beams are primarily responsible for regulating the linear mode dynamics, which contradicts the findings of Paper I. 
A similar conclusion can be drawn from the \emph{third panel}, where median growth rates of each group of modes are shown in solid lines. 
The beam induced modes seem to grow much faster---and therefore emit more power---than any other group by far in both moderately and mostly collisionally processed solar wind. 
To clarify this apparent contradiction, we plot the same median values in the \emph{bottom panel}, but normalized to their occurrence in each of the bins. 
This normalization clarifies that, even though the parallel beam induced modes grow quickly, they are not nearly as abundant as the core anisotropy IC mode, which is constantly present until the bulk of the core distributions become almost completely isotropic. 
As the core participation drops, the activity of $\alpha$ component, which generally has non-zero drift with respect to the core and therefore has ``slower collisional clock'' \citep{Kasper_2017,Alterman_2018_ApJ}, becomes more important. 
In parallel, the slowly growing oblique FM is constantly induced by the decrease in the Alfv{\'e}n velocity $v_A$ as the solar wind expands, and is apparent only when other free energy sources are depleted. 
Finally, collisionally old solar wind features FH modes that can easily arise in high-$\beta$ environments. 
They are likely induced by fluctuations of the VDF \citep{Verscharen_2016_ApJ,Arzamasskiy_2022_arXiv} and their median growth rates are very low, but this does not imply that their role in the solar wind evolution can always be neglected. 
\emph{Third panel} of Figure \ref{fig:power_nc_cba} also shows 80th percentile of each of the mode groups (dotted lines), which is between a factor of 3 and an order of magnitude above the median for each group, implying that even the modes that are generally weak can occasionally drive very intense plasma waves. 
We conclude this discussion by reminding the reader that the reasoning presented here is an overall insight that can provide a generalized description, but cannot be directly applied to isolated intervals and case studies. 
To access stability properties of any limited sample of VDFs, it is required to use either the \texttt{SAVIC} code presented here, or a traditional dispersion solver.  

%%%%%%%%%%%%%%%%%%%
%%% CONCLUSIONS %%%
%%%%%%%%%%%%%%%%%%%

%=====================================================================
\section{Conclusions}
\label{sec:conclusion}

After statistical assessment of the database of linear instabilities derived from the VDF fits of \emph{Helios} observations performed in Paper I, we continue our effort to provide a complete description of the behavior of solar wind instabilities.
Undertaking a detailed investigation of a phase space that spans over 20 variables, almost all of which can meaningfully impact the underlying physics, has turned out not to be feasible via traditional methods. 
In this intermediate installment of a series of articles on the topic, we managed to overcome the difficulty of handling the extensive multi-dimensional database by building a set of ML algorithms that can predict the VDF stability, estimate features of unstable modes, and classify them into groups defined by physical processes. 

We used our methods to investigate the overall participation of parallel and oblique modes driven by proton core, proton beam, and $\alpha$ VDF components to find that, although the parallel IC modes caused primarily by beam anisotropy emit the largest amount of power once they arise, their occurrence rate is not enough to make them the primary driver of the solar wind wave dynamics, except in the moderately collisionally old solar wind. 
In the young wind, the core induced IC instability is practically ubiquitous, while in the collisionally processed wind, close to LTE, the beam induced oblique and core Firehose are, for most intervals, the only remaining unstable modes. 

Improvements of the \texttt{SAVIC} code, including processing of new generic VDFs with \texttt{PLUME} as expanded training data sets as well as using observations from other spacecraft, including the \emph{Wind} database, as an additional training resource are planned for future work.
These improved versions will be incorporated in the publicly available code.

%Improvements of \texttt{SAVIC} code with generic data, processing of new generic VDFs with \texttt{PLUME}, and also using the \emph{Wind} database as an additional training resource is planned for subsequent versions of  \texttt{SAVIC}, but is outside of the scope of this paper. 

%Therefore, two objective of this paper was twofold. 
%First, in the follow-up article, we will use the tools developed here and in Paper I to provide a comprehensive model of solar wind instabilities evolution. 
%Second, application of the ML models developed for this purpose in any other stability related analysis done within the community can be very straight-forward, and for this purpose we provide the publicly available code and a comprehensive documentation on its contents and usage, while some examples and illustrations are given in Appendix \ref{ssec:guide}. 

%%%%%%%%%%%%%%%%%%%%%%%%
%%% ACKNOWLEDGEMENTS %%%
%%%%%%%%%%%%%%%%%%%%%%%%

\begin{acknowledgments}
M. M. Martinovi\'c and K. G. Klein were financially supported by NASA grants: 80NSSC22K1011, 80NSSC19K1390, 80NSSC23K0693, 80NSSC19K0829.
K.G.K. is supported by NASA ECIP Grant 80NSSC19K0912. 
An allocation of computer time from the UA Research Computing High Performance Computing at the University of Arizona is gratefully acknowledged.
The authors would also like to thank members of ISSI International Team \#563 supported by the International Space Science Institute (ISSI) in Bern for productive conversations regarding this work. 
\end{acknowledgments}

%%%%%%%%%%%%%%%%
%%% APPENDIX %%%
%%%%%%%%%%%%%%%%

%\appendix

%%%%%%%%%%%%%%%%%%
%%% REFERENCES %%%
%%%%%%%%%%%%%%%%%%

\bibliographystyle{aasjournal}
%\bibliography{Latex_Refs}

\end{document}